# Practical Evaluation of FFT-Based Thickness Extraction for Thick-Film Reflectometry

Zimu Zhou*[†], Enrique A. Lopez-Guerra[†], Michael Kwan, Chester Chien

*Western Digital, 5601 Great Oaks Pkwy, San Jose, CA 95119, USA*

*E-mail: zimu.zhou@wdc.com

[†]These authors contributed equally to this work

**Abstract**

Fast Fourier transform (FFT) is widely used for thick-film reflectometry because of its simplicity and computational efficiency. However, its performance under practical thick-film measurement conditions has received limited experimental evaluation. In this work, FFT, Linearized Reflectance Zero-Crossing (LRZ), and optical model fitting were compared using reflectometry measurements from a nominal 52 μm dielectric film acquired on a production wafer, with White-Light Interferometry (WLI) serving as an independent reference. Although FFT exhibited excellent repeatability, it showed systematic deviation from WLI (RMSE = 0.62 μm), whereas LRZ significantly improved accuracy (RMSE = 0.20 μm). Simulations and theoretical analysis indicate that the observed systematic deviation is consistent with spectral leakage caused by finite measurement windows and non-integer fringe periodicity. These results demonstrate that excellent repeatability does not necessarily imply high accuracy and highlight the advantages of zero-crossing-based approaches for thick-film thickness metrology.

Accurate thickness measurement of dielectric films is essential in semiconductor manufacturing, advanced packaging, optical coatings, and data-storage device fabrication[1). Reflectometry remains attractive for high-throughput in-line metrology because measurements can be performed rapidly without physical contact or destructive sample preparation. For thin films, thickness is commonly extracted using optical model fitting, where measured spectra are matched to theoretical reflectance models through iterative optimization[2). While highly accurate, model-based approaches require knowledge of optical constants, appropriate model selection, and computationally intensive fitting procedures.

For thicker films, interference fringes become increasingly dense and frequency-domain techniques based on FFT are frequently employed to estimate thickness directly from fringe periodicity[3),4). In this approach, the reflectance spectrum is transformed into wavenumber space and the dominant frequency component is converted into optical thickness. FFT-based methods are computationally efficient and require minimal prior knowledge of material properties, making them attractive for industrial applications. Although FFT has been successfully applied to thick-film reflectometry, practical production measurements are acquired over finite spectral windows, and the influence of this measurement condition on thickness accuracy has received comparatively little experimental attention.

Recently, model-free approaches based on fringe feature tracking have been proposed for thick-film reflectometry. The Linearized Reflectance Extrema (LRE)[5) method estimates thickness from extrema spacing, while the more recent LRZ[6),7) approach utilizes zero-crossings of a detrended spectrum to improve robustness in thick or partially absorbing films. Unlike FFT, these approaches estimate optical thickness directly from phase progression rather than from a Fourier peak.

In this work, FFT, LRZ, and optical model fitting were evaluated using measurements obtained from a nominal 52 μm dielectric film. While FFT-based fringe analysis has been widely reported in the optical metrology literature, comparatively few studies have evaluated its performance using production-scale wafer measurements referenced against an independent metrology technique. In manufacturing environments, measurement performance must be assessed not only in terms of theoretical frequency resolution, but also with respect to repeatability, systematic bias, and agreement with independent reference metrology. Therefore,

FFT, LRZ, and optical model fitting were evaluated using measurements acquired from a production wafer and benchmarked against WLI.

Commercial 200 mm wafer reflectometry (NANOSPEC® 9100 96 Series) and WLI (Bruker Contour GT-X) measurements were performed at 9 designated metrology test sites distributed across a production wafer with a nominal film thickness of approximately 52 μm. Thickness values were subsequently extracted from the reflectometry spectra using FFT analysis, LRZ analysis, and optical model fitting. The designated test sites were common to both measurement platforms to ensure direct comparison at identical wafer locations. WLI served as an independent reference technique because it employs a fundamentally different measurement principle from reflectometry and provides reliable thickness measurements for films in the thickness range investigated in this work[8]. Nine repeated measurements were acquired at each of nine designated wafer sites, yielding a total of 81 measurements which enables a statistically meaningful assessment of within-site repeatability, site-to-site variation, and agreement with the WLI reference measurements.

Figure 1 summarizes the site-average thickness measurements and within-site standard deviations obtained using each method. FFT exhibited excellent repeatability, with standard deviations generally below 0.04 μm across all measurement locations. LRZ demonstrated slightly larger variation, typically below 0.07 μm, while optical model fitting showed site-dependent repeatability ranging from nearly zero to approximately 0.12 μm. The FFT results initially appear attractive because of their excellent repeatability. However, repeatability alone does not necessarily guarantee measurement accuracy. Comparison against the WLI reference measurements reveals a substantially different picture.

Figure 2 compares thickness values obtained from each method against WLI. FFT exhibited the largest deviation from the reference measurements, producing a correlation coefficient of 0.884 and an RMSE of 0.621 μm. Residual analysis further revealed systematic offsets that were consistently larger than those observed for either LRZ or optical model fitting. In contrast, LRZ achieved a correlation coefficient of 0.985 with an RMSE of 0.204 μm. Optical model fitting produced the highest overall accuracy with $r = 0.997$ and $RMSE = 0.093$ μm.

The combination of Figs. 1 and 2 highlights an important distinction between repeatability and accuracy. Although FFT produced highly repeatable measurements, comparison with an independent WLI reference revealed a systematic offset. This suggests that under the measurement conditions investigated here, the dominant error source is associated with frequency estimation rather than measurement noise.

To better understand the experimental observations, it is convenient to express the reflectance spectrum in wavenumber space, $k_0 \equiv 2\pi/\lambda_0$. Under the two-beam approximation, the reflectance can be written as[5),9)]

$$R(k_0) = R_0 + R_1 \cos[\,2n_1 d_1\, k_0 - \phi_0\,] \quad (1)$$

Where

$$R_0 = r_{01}^2 + (1 - r_{01}^2)^2\, r_{12}^2$$
$$R_1 = 2|r_{01} r_{12}|(1 - r_{01}^2)$$

Here $R_0$ and $R_1$ are constants determined by the interface reflectivity and $\phi_0$ is a constant phase offset that absorbs the sign of $r_{01}r_{12}$. The oscillation frequency of Eq. (1) is directly proportional to the optical thickness $2n_1d_1$, which forms the basis of FFT-based thickness extraction. When the measurement window $[k_0^{\min}, k_0^{\max}]$ spans an integer number of oscillation periods of (1), the signal begins and ends at the same phase, satisfying the periodicity assumption inherent to the discrete Fourier transform. Under this condition, the signal frequency falls exactly on an FFT bin, and the spectral energy is concentrated at the frequency corresponding to the optical thickness $L = 2n_1d_1$, resulting in an unbiased thickness estimate. In practical measurements, however, the wavelength sweep is fixed by the instrument and rarely contains an integer number of fringe periods. Consequently, the measured signal is typically truncated at an arbitrary phase, introducing an artificial discontinuity at the window boundaries. One well-known consequence of finite measurement windows is spectral leakage, whereby energy spreads into neighboring frequency bins, causing broadening and shifting of the FFT peak[10)]. Although windowing functions can reduce sidelobes and zero-padding can improve apparent spectral resolution, neither approach eliminates the frequency bias arising from non-integer fringe periodicity and finite measurement windows[3)10)].

The practical implications of this behavior are illustrated in Fig. 3. Three simulated signals corresponding to identical optical thicknesses were generated, differing only in the number of fringe periods contained within the observation window. When the window contained an integer number of periods, the FFT peak coincided with the correct optical thickness. However, when the signal was truncated at non-integer fringe counts, the extracted FFT peak shifted substantially despite no change in the underlying film thickness. This behavior arises because FFT implicitly assumes periodic continuation of the measured signal. When the measurement window does not contain an integer number of fringe periods, discontinuities are introduced at the window boundaries, redistributing energy into neighboring frequency bins and shifting the apparent peak position.

As film thickness increases, more interference fringes are contained within a fixed spectral measurement window, making FFT peak localization increasingly sensitive to finite-window effects. In contrast, LRZ determines thickness from zero-crossing positions in the detrended spectrum[6)]. Because the method tracks fringe phase progression directly rather than relying on Fourier peak localization, it is less sensitive to the periodicity assumptions associated with FFT analysis. Furthermore, increasing fringe density provides additional zero-crossings for regression, improving statistical averaging rather than degrading accuracy.

In summary, FFT, LRZ, and optical model fitting were evaluated for thick-film reflectometry using WLI as an independent reference. FFT provided excellent repeatability across 81 production-wafer measurements and remains an attractive approach for rapid thick-film reflectometry. However, comparison with an independent WLI reference indicated systematic deviations under the measurement conditions investigated in this study. Simulations and theoretical analysis suggest that these deviations are consistent with spectral leakage associated with finite measurement windows and non-integer fringe periodicity. LRZ significantly reduced measurement error while preserving the advantages of a non-iterative and model-free workflow. These results suggest that phase-tracking approaches such as LRZ provide a practical alternative to FFT for thick-film thickness metrology, particularly when high accuracy is required without the complexity of optical model fitting.

**Acknowledgments**

This research did not receive any funding from external sources outside of Western Digital Corporation.

**References**

1) N. G. Orji, M. Badaroglu, B. M. Barnes, C. Beitia, B. D. Bunday, U. Celano, R. J. Kline, M. Neisser, Y. Obeng and A. E. Vladar, Nat. Electron. **1** [10], 532 (2018).

2) M. A. Taranov, B. G. Gorshkov, A. E. Alekseev, Yu. A. Konstantinov, A. T. Turov, F. L. Barkov, Z. Wang, Z. Zhao, M. S. D. Zan and E. V Kolesnichenko, Instruments and Experimental Techniques **66** [5], 713 (2023).

3) M. Quinten, SN Appl. Sci. [ DOI:10.1007/s42452-019-0866-9].

4) P. J. de Groot, Reports on Progress in Physics **82** [5], 056101 (2019).

5) Z. Zhou, E. A. Lopez-Guerra, B. Zhou, M. Kwan, P. Wilkens and C. Chien, Journal of Micro/Nanopatterning, Materials, and Metrology [ DOI:10.1117/1.JMM.24.2.024001].

6) Z. Zhou, E. A. Lopez-Guerra, I. Zana, V. Nguyen, N. Q. H. Tran, V. Huang, B. Zhou, G. Qian, M. Kwan, P. Wilkens and C. Chien, Metrology **6** [1], 13 (2026).

7) Z. Zhou, E. A. Lopez-Guerra, I. Zana, N. Q. H. Tran, B. Zhou, G. Qian, M. Kwan, P. Wilkens and C. Chien, in Metrology, Inspection, and Process Control XL, eds. H. Cramer and N. G. Schuch (SPIE, 2026) p. 82.

8) G. Huang, C. Cui, X. Lei, Q. Li, S. Yan, X. Li and G. Wang, *Micromachines (Basel).*, 2025, 16.

9) H. Angus Macleod, Thin-Film Optical Filters, Fifth Edition (CRC Press, 2017).

10) F. J. Harris, Proceedings of the IEEE **66** [1], 51 (1978).

## Figures

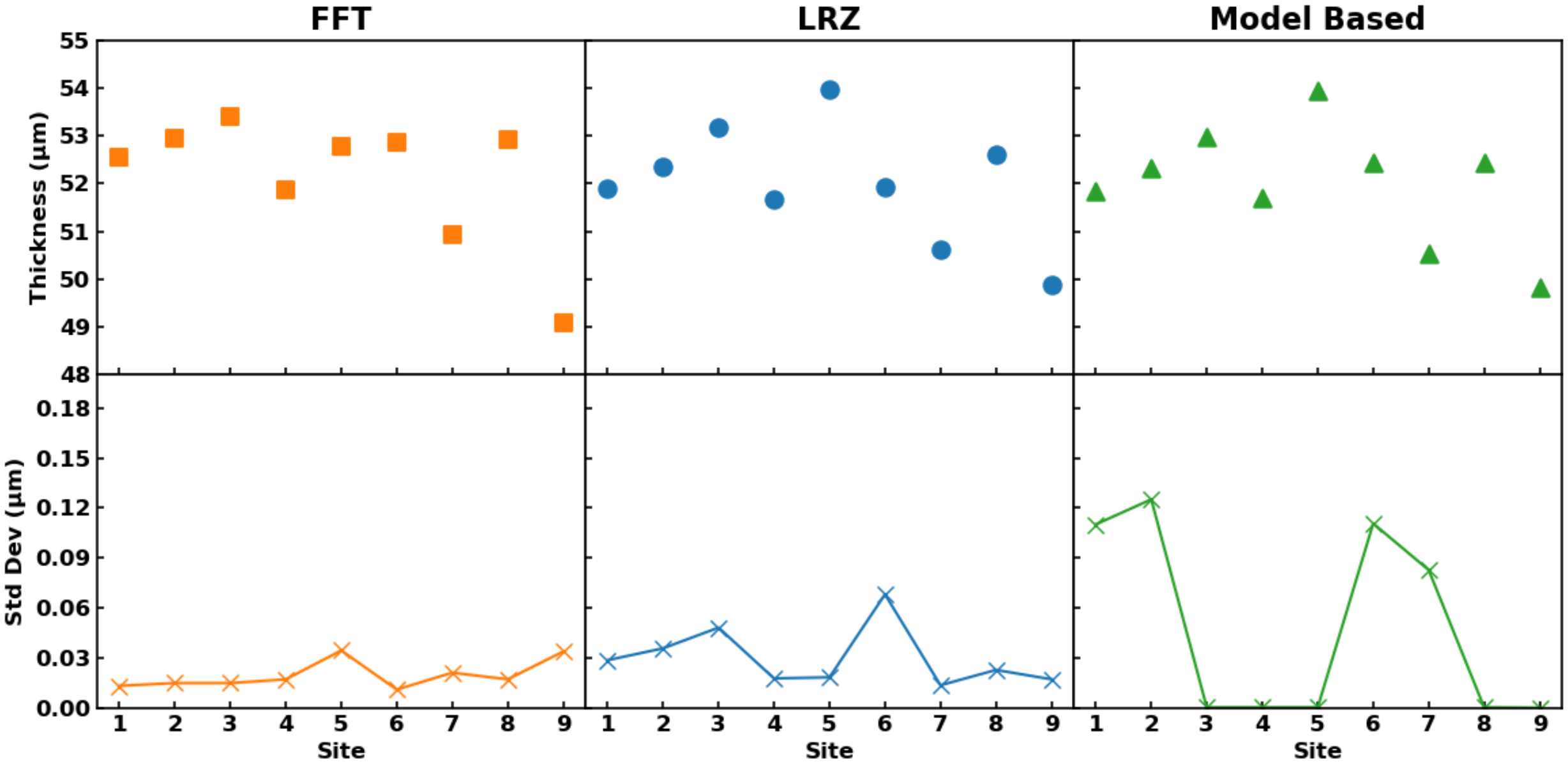

**Fig. 1.** Thickness measurements obtained from FFT, LRZ, and optical model fitting across nine wafer sites (top row), together with the corresponding standard deviation from nine repeated measurements at each site (bottom row). All three methods exhibit good repeatability, with FFT showing the lowest measurement variation.

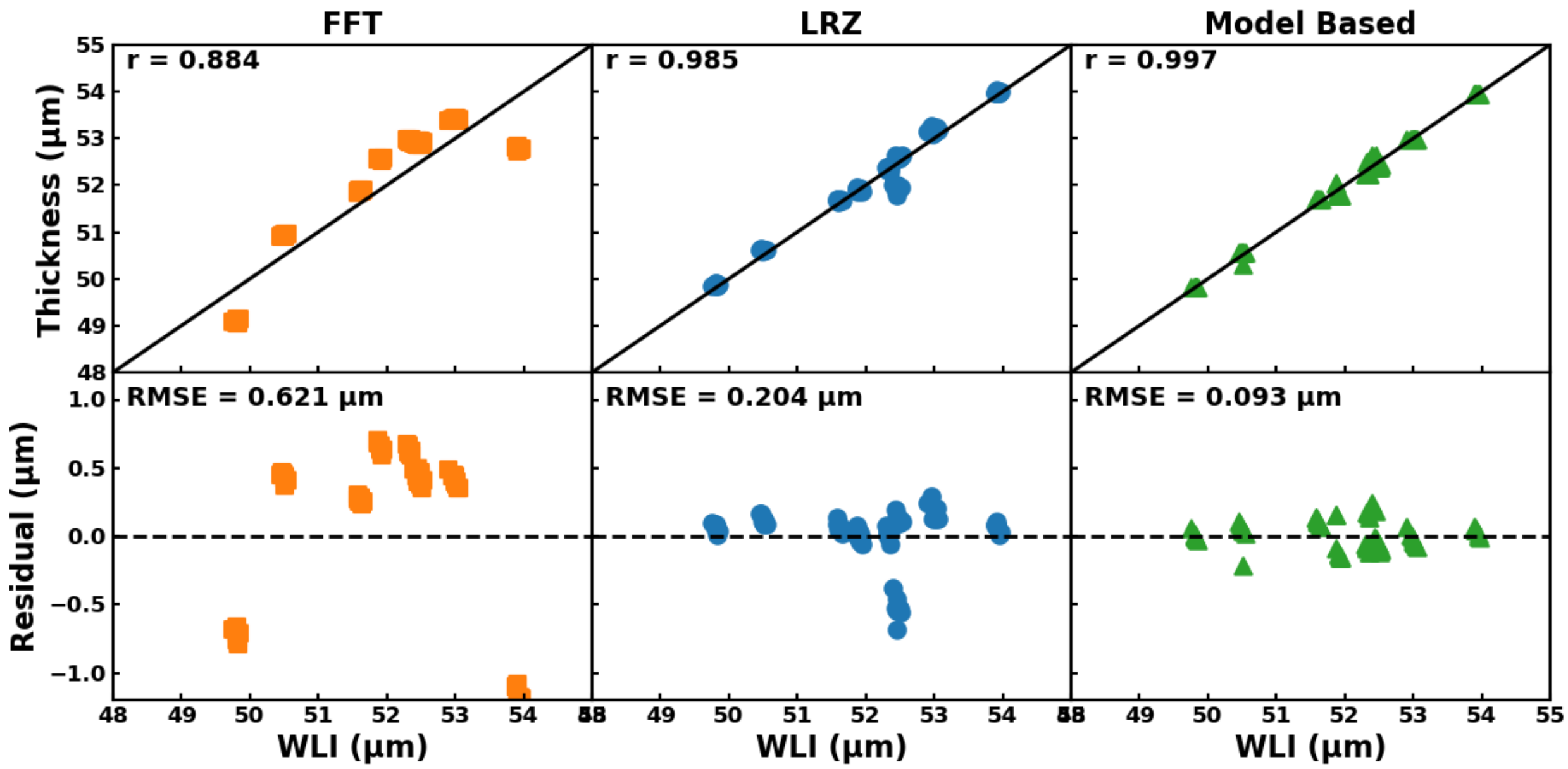

**Fig. 2.** Comparison of thickness measurements obtained from FFT, LRZ, and optical model fitting against the WLI reference. The upper panels show site-average thickness values, while the lower panels show residuals relative to WLI. FFT exhibited the largest systematic deviation (RMSE = 0.621 µm, r=0.884), despite its excellent repeatability shown in Fig. 1. LRZ substantially improved agreement with WLI (RMSE = 0.204 µm, r=0.985), while optical model fitting achieved the highest accuracy (RMSE = 0.093 µm, r=0.997).

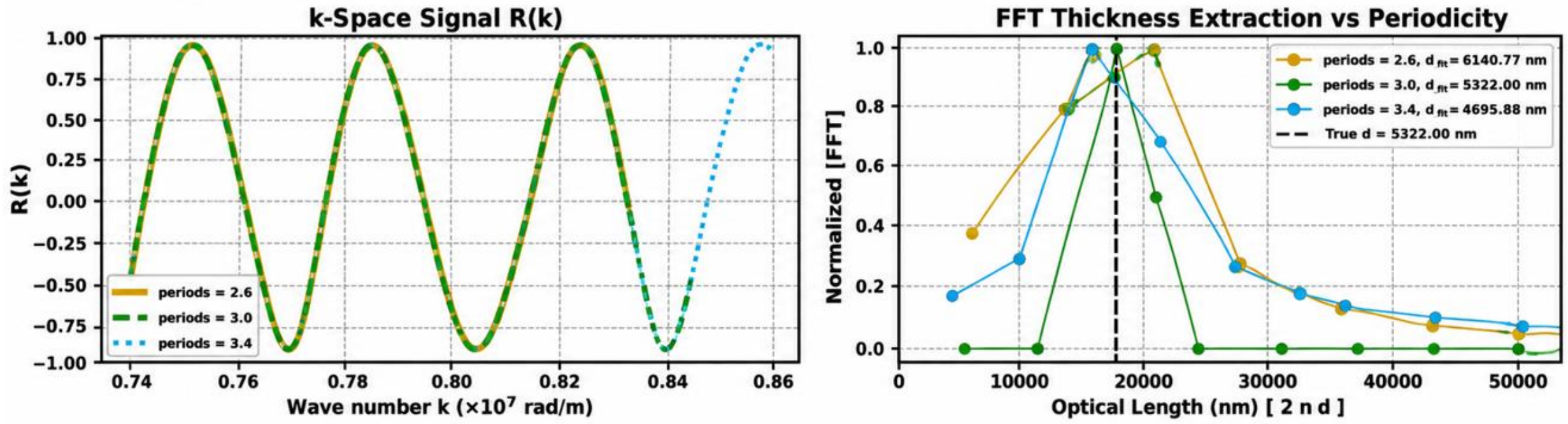

**Fig. 3.** Simulated demonstration of spectral leakage in FFT-based thickness extraction. Signals with identical optical thicknesses but different numbers of fringe periods were analyzed using FFT. Integer-period truncation (3.0 periods) produces an unbiased FFT peak, whereas non-integer truncation (2.6 and 3.4 periods) causes peak shifting and thickness estimation error. The results illustrate how finite measurement windows can introduce systematic bias in FFT-based thick-film metrology